\documentclass[12pt]{article}
\usepackage{natbib}
\usepackage{graphicx}
\usepackage{epsfig}
\usepackage{amssymb}

\begin{document}

\baselineskip24pt

\centerline{\bf  Chronology and Sources of Lunar Impact Bombardment} 

\bigskip
\centerline{Matija \' Cuk$^{1, 2}$}

\bigskip

\centerline{$^1$Smithsonian Astrophysical Observatory}
\centerline{60 Garden Street, Cambridge, Massachusetts 02138}

\centerline{$^2$Department of Earth and Planetary Sciences, Harvard University}
\centerline{20 Oxford Street, Cambridge, Massachusetts 02138}

\bigskip

\centerline{E-mail: cuk@eps.harvard.edu}

\vspace{24pt}
\centerline{Resubmitted to Icarus}
\centerline{November 26$^{\rm th}$ 2011.}

\vspace{24pt}

\centerline{Manuscript Pages: 55}

\centerline{Tables: 2}

\centerline{Figures: 6}

\newpage

Proposed Running Head: Lunar Impact Bombardment

\vspace{48pt}

Editorial Correspondence to:

Matija \' Cuk

Carl Sagan Center

Seti Institute

190 N Bernardo Ave

Mountain View, CA 94043

Phone: 650-810-0210

Fax: 650-961-7099

E-mail: cuk@eps.harvard.edu

\newpage

ABSTRACT: The Moon has suffered intense impact bombardment ending at 3.9 Gyr ago, and this bombardment probably affected all of the inner Solar System. Basin magnetization signatures and lunar crater size-distributions indicate that the last episode of bombardment at about 3.85 Gyr ago was less extensive than previously thought. We explore the contribution of the primordial Mars-crosser population to early lunar bombardment. We find that Mars-crosser population initially decays with a 80-Myr half-life, with the long tail of survivors clustering on temporarily non-Mars-crossing orbits between 1.8 and 2 AU. These survivors decay with half-life of about 600 Myr and are progenitors of the extant Hungaria asteroid group in the same region. We estimate the primordial Mars-crosser population contained about 0.01-0.02 Earth masses. Such initial population is consistent with no lunar basins forming after 3.8 Gya and the amount of mass in the Hungaria group. As they survive longer and in greater numbers than other primordial populations, Mars-crossers are the best candidate for forming the majority of lunar craters and basins, including most of the Nectarian system. However, this remnant population cannot produce Imbrium and Orientale basins, which formed too late and are too large to be part of a smooth bombardment. We propose that the Imbrian basins and craters formed in a discrete event, consistent with the basin magnetization signatures and crater size-distributions. This late "impactor shower" would be triggered by a collisional disruption of a Vesta-sized body from this primordial Mars-crossing population \citep{wet75} that was still comparable to the present-day asteroid belt a 3.9 Gya. This tidal disruption lead to a short-lived spike in bombardment by non-chondritic impactors with a non-asteroidal size-frequency distribution, in agreement with available evidence. This body ("Wetherill's object") also uniquely matches the constraints for the parent body of mesosiderite meteorites. We propose that the present-day sources of mesosiderites are multi-km-sized asteroids residing in the Hungaria group, that have been implanted there soon after the original disruption of their parent 3.9 Gyr ago.


\newpage

\section{Introduction}

Small bodies in the Solar System are usually divided into stable populations (like the main asteroid belt, Jupiter Trojans or the Kuiper belt) and the unstable, planet-crossing populations (like the near-Earth asteroids, comets or the Centaurs). Stable populations are seen as fossilized remnants from the formation of the Solar System, while the planet-crossing small bodies are short-lived and are constantly replenished from these stable reservoirs. There are some exceptions to this classification, like the Trans-Neptunian Scattered Disk, which is very long-lived but evolves appreciably over the age of the Solar System \citep{vol08}. The small Hungaria asteroid group, on low-eccentricity, high-inclination orbits just beyond Mars, has also been suggested as a constantly eroding population \citep{mce10}, but the full implications of the Hungarias' long-term decline have not been explored yet.

The issue of small body stability and instability is important for determining the impact history of planets and satellites. The Moon is the best studied body in the Solar System after Earth, and reveals a record of the system's early history that has since been erased on our planet. The lunar samples returned by the Apollo missions firmly established that the Moon was subject to intense bombardment at 3.85 Gyr ago (Gya) \citep{ter74}. The nature of lunar bombardment before that time is controversial, as is the source of the impactors at 3.85 Gya. The terms "Lunar cataclysm" (LC) and "Late Heavy Bombardment" (LHB) have been introduced to describe this bombardment, and the two terms are now used almost interchangeably. A non-primordial source of the lunar cataclysm at 3.85 Gya is based on two lines of argument. One argument is that the decay of lunar bombardment at the time of the large Imbrium impact at about 3.85 Gya is too rapid to be consistent with a slowly decaying primordial population \citet{ccg07}. Independently, dynamical calculations show that the primordial small bodies should have been all but extinct by 3.85 Gya \citep{bot07}. This leads to a conclusion that the lunar cataclysm impactor population must have become unstable after a period of stability. 

The only plausible mechanism for a delayed instability of a numerous small body population would be a late rearrangements of the planets. This could be a local event, involving an additional terrestrial planet \citep{cha07}, or global, affecting the giant planets and, indirectly, the rest of the system \citep{lev01, tsi05, gom05}. While there is independent evidence for an unstable episode in the Solar System's history \citep{mor05, nes07, mor09, mor10}, there is no independent evidence about the timing of this instability; it is most often assumed to coincide with the lunar cataclysm \citep{gom05}. 

In \citet{cuk10} we argue that the record of the cataclysm does not match the predictions of the planetary migration-based models. Youngest basins (Imbrium and Orientale) seem to have formed much later than the older basins (see section 3), and were produced by a non-asteroidal impactor population \citep{cuk10}, while planetary migration predicts the late impactors to be derived from main-belt asteroids \citep{gom05}. More generally, the necessity of a connection between the planetary instability and bombardment at 3.85 Gya hinges on the lack of other plausible explanations for the late delivery of lunar impactors. In this paper, we will show that there should still be a significant number of primordial Mars-crossers present at 3.85 Gya, so that this population can indirectly trigger a relatively short-lived lunar cataclysm, in full agreement with the available evidence.

\section{Lunar Chronology: Basin Magnetization}

The Moon, unlike Earth, currently does not have a global magnetic field. However, lunar surface does exhibit magnetization and these local variations in the magnitude and direction of local crustal fields have been mapped by instruments flown on Lunar Prospector spacecraft \citep{hal01, hoo01}. The correlation between the crustal magnetization and other surface features (basins in particular) has been controversial and is certainly complex \citep{hal03}. Some of the magnetic anomalies may actually be antipodal to basins \citep{mit08} and therefore likely formed by a transient plasma cloud \citep[][ this way younger impacts can even re-magnetize older basins] {hoo08}, but this is not accepted by all researchers \citep{wie11}. Here we are interested in lunar basin magnetization because it can offer us an alternative way of classifying basins that can help determine the  chronology of their formation. Magnetic anomalies inside basins are generally weaker than those outside basins, but are more likely to be of thermoremanent origin and therefore record a past steady magnetizing field (Lon Hood, personal communication).

\citet{moh08} find that the Imbrian basins (as well as the late Nectarian basin Hertzsprung) have very different magnetization properties from older basins. While the sources of ancient lunar magnetism are unknown, \citet{moh08} pointed out that there appears to be a correlation between the basins' remanent magnetization \citep{mit08} and their relative age \citep[cf. ][]{hal03}. Most interestingly, only the youngest few basins have weak or no remanent magnetization. In contrast, some Nectarian basins, especially Crisium, have magnetic signatures that are fully consistent with a global dynamo-generated dipole field \citep{hoo11}. More recent analyses of Lunar Prospector data \citep{pur10} produced better maps that strengthen the dichotomy between the stronger magnetic anomalies within most Nectarian basins and much weaker magnetization of Hertzsprung and Imbrian basins. Still, the conclusions of \citet{moh08} are based on spherical harmonic model maps \citep[such as ][]{pur10} and will need to be confirmed using direct mapping methods \citep{hoo11}.

The simplest explanation for these differences in basin magnetization is that the lunar magnetic field was absent at the time of formation of basins younger than Hertzsprung, but present before that. This hypothesis has profound implications for basin formation chronology. If there was no global magnetic field at the time of the formation of Imbrium, all other basins that formed during the same short-lived lunar cataclysm spike should record the same (weak or zero) remanent magnetization. Therefore, only Imbrium, Orientale, Schr\" odinger, Hertzsprung and (possibly) Mendeleev basins have magnetic signatures and relative ages consistent with a nearly co-eval formation in a spike. We note that this interpretation of magnetic data does not require Hertzsprung and Mendeleev to be result of the cataclysm as the field may have died long before the impact spike, and they could have formed between the end of the internal field and the lunar cataclysm.

Our hypothesis has an important implication for the age of the Serenitatis basin. Since Serenitatis basin does record magnetic fields \cite[][ Lon~Hood, personal communication]{moh08}, our scenario places its formation significantly before the 3.9~Gya age of Apollo 17 samples. This is in line with interpretation of \citet{has98} and \citet{spu11} who argue that those samples originated in the Imbrium event, in contrast with the conventional interpretation that they come from Serenitatis \citep{wil87}. Therefore, Imbrium provenance of 3.9~Gya old Apollo 17 samples would support our hypothesis, while the 3.9 Gya age for Serenitatis would contradict our interpretation of lunar chronology.



One could argue that largest impacts could have affected the heat flow in the lunar mantle and effectively shut down the dynamo, making the end of dynamo synchronous with the impact spike \citep{ark10}. However, lack of magnetic signature inside of the pre-Imbrium Hertzsprung basin indicates that the decay of lunar magnetic field preceded the giant Imbrium impact, while magnetic anomalies within the Crisium and Serenitatis basins rule out shutdown by the those large upper Nectarian impacts. Hertzsprung and Mendeleev basins were probably the first to form after the end of dynamo, but they are relatively small and therefore poor candidates for an end-dynamo event.

Magnetization of lunar surface sampled by Apollo missions rocks also carries information about the past lunar magnetic field. \citet{law08} find that there is no firm evidence of lunar dynamo in (mostly Imbrian-age) Apollo samples, despite previous claims, while \citet{gar09} identify traces of a likely lunar dynamo in 4.2 Gyr-old sample. Some researchers find evidence for a dynamo as late as 3.7 Gya, recorded in mare rocks \citep{she10}, but it is not clear if those results could be reconciled with Lunar Prospector data. In conclusion, we propose that the lunar basins coeval with Imbrium basin most likely all lack strong thermal remanent magnetization, and that this property can be used to deduce the chronology of basins not sampled by the Apollo landings.

\section{Lunar Chronology: Impactor Size Distributions}

Crater counting is the basic technique for determining relative ages of planetary surfaces, with the older surfaces having accumulated more craters. Using absolute crater counts of superposed craters, basins can be grouped into early Nectarian (like Nectaris or Moscoviense), late Nectarian (among others, Serenitatis, Crisium, Mendeleev and Hertzsprung) and Imbrian (Imbrium, Orientale and Schr\" odinger).  Given that late-Nectarian basins cluster together in crater densities \citep{wil87}, and the limited resolution of crater counting, it is impossible to securely put all late Nectarian basins in a chronological order. While \citet{wil87} consider Hertzsprung to be the youngest Nectarian basin, \citet{fas11} identify Mendeleev as being the least cratered. Direct superposition of basin materials can help but is not available for all basins (Mendeleev and Hertzsprung do not overlap with any other late Nectarian basins). However, it may be possible to compare the size-frequency distributions of of craters that are thought to have formed before and after the decay of the lunar magnetic field.

\citet{cuk10} point out that most of the morphologically young ''class 1'' craters \citep{str05} must be of lower Imbrian age. If the ''class 1'' crater size-distribution is representative of the Imbrian impactor population, then the Imbrian impactors has a SFD very different from the preceding impactor populations recorded on the heavily cratered highlands \citep{str05, cuk10}. Separate measurements not dependent on morphological classification would help confirm or refute this distinction between Imbrian and pre-Imbrian impactors. Ideally, these would be counts of all craters superposed on an Imbrian surface unit. \citet{str77} report counts of craters superposed on the Orientale basin ejecta blanket, and state that it is consistent with the SFD of craters on lunar highlands \citep[later called "population 1" by ][]{str05}. Using the Kolmogorov-Smirnov (KS) test we find that the \citet{str05} highland and class 1 populations are incompatible at 0.1\% level, but neither can be distinguished from the \citet{str77} post-Orientale craters at 10\% level. Unfortunately, it appears that about 200 craters identified by \citet{str77} on the Orientale ejecta blanket are not sufficient to clarify the Imbrian impactor distribution, especially as the published counts are binned. This weak statistical significance of the Orientale basin "population 1" crater SFD claimed by \citet{str77} is still disputed by some researchers \citep{mal11, cuk11}

\citet{wil78} classified craters based on their stratigraphic relationships. They find that Imbrian (and younger) craters had a different SFD than those formed in the Nectarian period. Table \ref{KStest} summarizes the comparison of \citet{wil78} Nectarian and combined Imbrian, Eratosthenian and Copernican ("Imbrian and younger") crater size distributions to \citet{str05} highland and class 1 craters (for this comparison we used only craters with $D>22.6$). Using only these overlapping subsets of those crater populations reduces statistical significance, but we can still conclude, based on KS tests that Imbrian (and younger) craters are incompatible with the highland distribution (at $<$0.5\%) , and compatible with the class 1 craters ($>$10\%). On the other hand, Nectarian craters are compatible to highland population ($>$10\%) and much less similar to the class 1 set (at $<2.5\%$; significance is low as there are only $<200$ class 1 craters with D$>22.6$km).

\begin{figure*}
\includegraphics[angle=270, scale=.8]{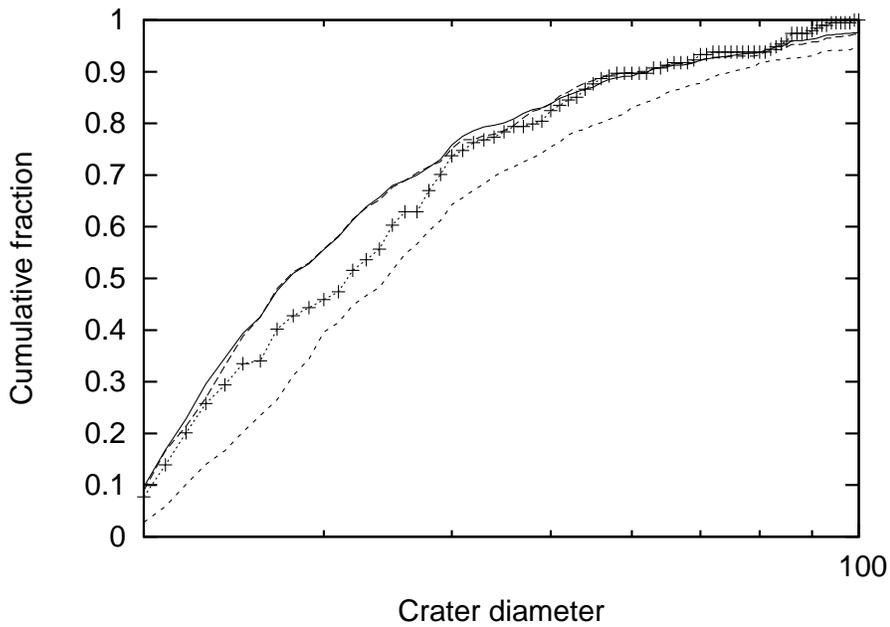}  
\caption{Comparison between the cumulative distribution of different populations of lunar craters with $D>$20~km. The (very similar) solid and long dashed lines plot the crater distributions from \citet{wil78}: Imbrian (solid), Imbrian and younger (long dashes), and Nectarian (short dashes). The intermediate cross-dot line plots the combined distribution of craters superposed on Hertzsprung, Imbrium, Orientale and Schr\" odinger basins ("HIOS" basins). }
\label{20km}
\end{figure*}

Finally, \citet{wil87} report counts of D$>$20~km craters superposed on a number of large lunar basins, which are used to determine the basins' relative chronology. Unfortunately, almost all basins have less than hundred superposed craters of that size, making statistically robust SFD comparisons impossible. Since the magnetization data imply that Hertzsprung and younger basins may have formed in a separate event from the preceding basins, we combined the crater counts on Hertzsprung, Imbrium, Orientale and Schr{\" o}dinger basins ("HIOS") and plotted them in Fig. \ref{20km} alongside complete Imbrian (and younger) and Nectarian crater populations from \citet{wil78}. While the HIOS craters are incompatible with the Nectarian set ($<$1\%; Table \ref{KStest}), they are more compatible with the Imbrian and younger craters ($2.5\% < p < 5\%$). Additionally, \citet{wil78} Nectarian and Imbrian and younger craters are incompatible with each other at 0.1\% (as are Nectarian and Imbrian proper), confirming once more that they represent different populations. 

Therefore, both morphological and stratigraphic classifications of craters agree that there were two different impactor populations in the Nectarian and Imbrian periods. Nectarian period was dominated by "population 1" impactors \citep[to use the nomenclature of ][]{str05}, which also seems to dominate the total population of lunar highland craters, and has SFD similar to that of main belt asteroids in the relevant size range \citep{str05}. However, Imbrian impactors ("population 2"), also represented by \citet{str05} "class 1" craters, have a different SFD that is less "top heavy" and closer to having the cumulative exponent -2 \citep{cuk10}. Comparisons of these populations to the "ground truth" of craters that likely formed after the decay of lunar magnetic field ("HIOS" set) favor the population 2 impactors over population 1 ones, despite problems with relatively small number statistics. 


In summary, both basin magnetization signatures and cratering data agree with a transition between two impactor populations at about the stratigraphic level of the Hertzsprung basin, implying that there was something different about the way the Imbrian basins (together with possibly Hertzsprung) formed. In the next section we will explore possible dynamical sources of these two impactor populations.

\section{Mars-Crosser Dynamics}

Mars-crossers are small bodies that cross the orbit of Mars, but not those of Earth or Jupiter. While Amors are technically Mars-crossers, their perihelia ($q < 1.2$~AU) practically couple them to Earth and their dynamics is similar to that of other Near-Earth Asteroids. Also, bodies with $a > 2.5$~AU or in resonances with Jupiter evolve much faster than the population we are interested in, and will not be discussed here.  

Mars-crossers are interesting because, while unstable, they have lifetimes significantly longer than other planet-crossing populations in the inner Solar System. This is simply a consequence of Mars's small size, making it the most ineffective among planets at clearing its orbit \citep{sot06}. If we are interested in the lunar bombardment at 3.9 Gya, primordial Mars-crossers are clearly a population of interest \citep[cf. ][]{wet75}, as a significant portion of them may have survived for 600 Myr. The most comprehensive exploration so far of the contribution of primordial planetesimals to the LC/LHB was done by \citet{bot07}, who found that the primordial inner solar system planetesimals could not have formed the late Imbrium and Orientale impact basins, which undoubtedly formed after 3.9 Gya. However, \citet{bot07} considered only bodies on high-inclination orbits with perihelia below 1.2~AU, excluding Mars-crossers. Therefore, their results need to be reevaluated by including a primordial Mars-crossing population.

We study the fate of Mars crossers by numerically integrating an initial grid of 1000 particles using a symplectic integrator SWIFT-rmvs3 \citep{lev94}. The test particles have been placed in an semimajor-axis and inclination grid, uniformly filling the box with sides of $1.75-2$~AU and $0-20^{\circ}$. All particles had $e=0.2$, resulting in initial perihelia in the $1.4 < q < 1.6$~AU range. The eight planets were started on their present orbits and with their current masses and sizes, and the timestep was 5 days. The integrations were completed on Harvard University's "Odyssey" computing cluster.

Figure 2 shows the distribution of surviving particles after 300 Myr. At that time, 87 of the original thousand were still in the system. Most of the particles are widely scattered through the Mars-crossing region, while some are Earth and Venus-crossers. At this point we cloned the remaining bodies by replacing each of them by 20 new particles with identical orbital elements, save for slightly different semimajor axis. We then continue integrating these 1740 particles until the time of 2 Gyr, when fewer than 30 particles remain, and the results cease being statistically significant. Figure 3 shows that population at 1 Gyr, when there are still 105 particles left. The surviving particles cluster in the region just beyond Mars with inclinations of 15-30$^{\circ}$ and low eccentricities. Figure 4 compares this population to present-day Hungaria asteroids. We find that the core of this late surviving Mars-crosser population matches the Hungaria group rather well.

Figure \ref{tail} shows the decay of our simulated population over time. The decay can best be described as a transition between two exponential decays. At first test-particles decay with a half-life of about 80 Myr. This is similar to findings of \citet{mor01} and the long tail found by \citet{bot07}. After about 600 Myr a slower decay predominates, with a half-life of about 600 Myr. This is in agreement with the lifetime of large Hungarias (mostly unaffected by radiation forces) found by \citet{mce10} and similar to the results of \citet{mil10}. 

While the overall decay of Mars-crossers and their evolution into proto-Hungarias appears orderly, it must be noted that Fig. \ref{tail} shows the collective behavior of particles across a range of simulations. SWIFT-rmvs3 integrator treats the test particles as massless, but they do have a subtle effect on planetary orbits, as the behavior of test particles triggers timestep reduction during planetary encounters. Using different timesteps at different times will make the simulations diverge slowly even if the planetary initial conditions were the same. This feature of SWIFT-rmvs3 is in this case desirable, as it allows us to show evolution of particles in ``parallel universes" and minimizes the effects of atypical planetary evolution in any one simulation. However, one must keep in mind that the Solar System had a singular history and if we were somehow able to repeat our experiment using the system's true history, there were likely deviations from the smooth decay shown in Fig 4. Most important departures from the average decay rates are likely to happen while the long-term maximum eccentricity of Mars is growing or shrinking \citep{las94, las08}, which would destabilize previously non-Mars-crossing bodies, or decouple some Mars-crossers from the planet, respectively. 


Two other limitations of this experiment need to be noted: we start with the current Solar System, and ignore General Relativity. There could be substantial differences between Fig 4 and the real decay rate of Mars-crossers if Mars started out of a less (or more) eccentric orbit than it has now \citep{cuk08, bot10}. The exclusion of Relativity mainly causes Mercury to be less stable than it is in the real Solar System \citep{bat08, las08}, and we had to throw out a number of simulations due to the inner planets becoming unstable. In those cases, we re-scale the results for the remaining population at the next benchmark, ignoring all losses to planetary instability.


\section{Collisional Evolution of Mars-Crosser Population}

Our numerical experiment utilizes massless test particles, so it cannot give us a direct measure of the amount of mass involved. To estimate the possible mass of this Mars-crossing population, we need to compare our population of test particles to lunar impactors at some point of time. For an estimate of the mass in these particles, we need to know the size-distribution of bodies making up this population. Here we will assume that the primordial Mars-crossers had the same size distribution as the asteroids in the inner main belt (interior to 2.5~AU). In this range, there are about 150 asteroids larger than 22 km (which would make a $D>$300~km basin on the Moon), about 50 bodies larger than 60~km (that could make a $D>900$~km basin) and a single Vesta with $D>$500~km. Since about 7-8\% of all particles eventually hit Earth (Table \ref{fates}), and about 5\% of that number should hit the Moon, we estimate that one in about 300 of our particles hit the Moon. Since no basins formed after 3.8 Gya \citep{ccg07}, there should have been no more than about 400 22-km-sized bodies left among Mars-crossers at 700 Myr. As there were about 200 Mars-crossers left at this point in our simulation, we estimate that each of our second-generation particles correspond to two planetesimals of about 22~km in diameter. 

One important reality check of this estimate is to compare the present Hungarias to our test particles. Starting with about 100 test particles at 3.5~Gyr ago and a half-life of 600 Myr, we expect about four 22-km Hungarias to survive to the present epoch, ignoring collisions and radiation forces. Hungaria region presently contains the remnants of one body of about that size that was disrupted about 0.5~Gya \citep{war09}, with a likely another, somewhat older family present \citep{wil92}. Therefore we seem to produce too many Hungarias by less than a factor of two, a relatively small disrepancy given the number of assumptions. While ignoring the Yarkovsky effect for bodies this size is likely justified \citep{mce10}, we do not generally expect asteroids smaller than 50~km to survive intact against collisions over the age of the Solar System \citep{min09}, so we need to address the collisional evolution directly.  

Hungarias experience significant bombardment by inner main belt asteroids, especially Flora family members \citep{war09}, so the fact that recent families are seen likely means that two 22~km body should have survived the collisional attrition almost down to the present. Using approach of \citet{wet67}, we calculate that the inner main belt asteroids have a probability of about 2.7$\times10^{-18} {\rm km^{-2} yr^{-1}}$ of striking Hungarias, and that the average collisional velocity is about $10$~km/s. Therefore a body of only $0.4$ km across should be able to disrupt a 22-km Hungaria. If there are about $2 \times 10^5$ such bodies in the inner main belt, we get about 3-Gyr half-life. Flora family has a 4 times higher collisional probability \citep{war09}, while containing about 20\% of the inner belt's members, increasing the number of impacts by about a factor of two for the last Gyr or so \citep{nes02}. Therefore our rough estimates are in line with the 22-km Hungaria population decaying by a factor of two through collisions between the end of the Mars-crosser collisional equilibrium and the present epoch. This is separate from the 100-fold dynamical depletion that is independent of asteroid size. We conclude that our estimate of the Mars-crosser population based on the cessation of basin formation at 3.8 Gya is roughly consistent with the observed population of Hungarias. 

Extending this estimate to our initial conditions, we find that the original population of Mars-crossers was 400 times larger than our comparison population (asteroids with $2<a<2.5$), and had a mass 60 times that of the present asteroid belt (or 3\% of Earth's mass). This mass is rather large and probably close to the upper limit of what Mars could have scattered, but it is still a small fraction of the mass in terrestrial planets. While much more numerous than main-belt asteroids, our Mars-crossers are all interior to the sweep of the $\nu_6$ secular resonance, which may have depleted the asteroid belt \citep[][as we explain the lunar cataclysm by other means, we assume that the planetary migration and $nu_6$ sweeping happened early on]{min09}. Alternatively, if there was a truncation in the inner system planetesimal disk \citep{han09, wal11}, we would expect a steep decline of the density of planetesimals as we move from the terrestrial planet zone to the present asteroid belt. In any case, the survival of present Hungaria asteroids require an initial non-Mars-crossing population interior to 2~AU ("proto-Hungarias") similar to that of the inner belt, as current replenishment mechanisms are not very efficient \citep{mce10}. Unless these low-$e$ Hungarias were planted by some unknown mechanism that specifically produces low-$e$ (but high-$i$) orbits, they had to be captured from among the Mars-crossers. As we find the capture efficiency to be on the order of a percent (Fig \ref{tail}), it seems that we cannot avoid having hundreds "Vestas" on Mars-crossing orbits early in the history of the solar system. 

Note that this population initially has about $10^4$ bodies that can disrupt Vesta, but their half-life is only about 80 Myr, so the probability of Vesta being disrupted is only 20\%, so the survival of Vesta is not by itself in conflict with such a large primordial Mars-crosser population. 

There is at least one serious caveat to this estimate. Our initial conditions are rather arbitrary and we start the integration with the already dynamically "hot" population, which may not be correct interior to the $\nu_6$ sweep. While it is likely that a cold population would eventually get heated and end up evolving similarly to our sample, orbital distribution of the initial population has profound implications for collisional evolution. Since our population monotonically decays, most intense collisional grinding happens early on when the Mars-crossing region is most densely populated. In the following sections we discuss the number of surviving Vesta-sized bodies at later epochs, but most bodies of that size would be disrupted early on if the initial population was indeed dynamically "hot" like our initial conditions. So while our simulation can be used as a proof that a generic Mars-crossing population decays with a half-life of 80 Myr and about 1\% of it enters the Hungaria region, the simulation plotted in Figs. 2-5 cannot fully correspond to the real history of the Solar System.

A more natural setup would be a significantly "colder" (and probably even larger) disk beyond Mars that Mars scatters and may even migrate into \citep[cf.][]{min11}. This kind of interaction requires using massive planetesimals and is outside of the scope of this work. Therefore our calculation for what happens before 4.3-4.4 Gya are not to be taken literally, and likely only the evolution of Mars crossers after that time is reasonably well described by our simulations. This caveat does not affect our conclusion that Hungarias are a natural long tail of Mars crossers, and that the extant Hungarias indicate a large but reasonable primordial population in the region between Mars and the asteroid belt.


\section{Lunar Bombardment by Mars Crossers}

After we have established an estimate of the Mars-crosser flux, we can compare it to the lunar record. If the last 300~km basin formed no later than 3.8~Gyr, when did the last 900+~km basin form? Using Fig. 4, the most likely time for this would be about 4.0~Gya. One could argue that the formation of Imbrium at 3.85-3.87~Gya was a plausible fluke, but Imbrium was shortly followed by 900-km Orientale impact \citep{sto00}. While we avoid making a case based on small number statistics, we note that this top-heavy Imbrian basin distribution inspired George Wetherill to look for alternatives to asteroid impacts as a source for the lunar cataclysm \citep{wet75}. When we compare our results with those of \citet{bot07}, we find that the formation of Imbrium and Orientale \citep[within the time limits set by][]{bot07} is a 1.5-sigma ($<$10\%) event in our model, significantly more likely in our model than \citet{bot07} found using NEA-like initial conditions.

There is, however, a strong argument against Imbrium and Orientale simply being relatively low-probability events that happened unexpectedly late. Apart from the basins, Imbrian system on the Moon contains a large number of smaller craters. Using numbers of $D>$20~km craters from \citet{wil78}, we find that at least 20 such craters formed per $10^6$~km$^2$ in the Imbrian period (together with three basins), compared to about 50~per~$10^6$~km$^2$ (and 10-12 basins) in the preceding Nectarian period. Therefore the two largest Imbrian impactors were accompanied by (at least) a proportionate number of small bodies, and could not have been statistical flukes drawn from the same Mars-crossing impactor distribution that formed the Nectarian basins. 

It is thus very unlikely that the primordial Mars-crossers could be the chief source of Imbrian impactors. Could they be the main source of Nectarian bombardment? The size-distribution of Nectarian impactors is certainly in agreement with the size-frequency distribution of main-belt asteroids \citep{wil78, str05}, which is also our assumed size-distribution of Mars crossers, which were in collisional equilibrium. While we do not know the actual SFD of primordial Mars-crossers, there is no reason to believe it would be different from that of neighboring main belt asteroids. And after 4.4-4.3 Gya they would have a significantly higher flux than impactors derived from Earth-crossing planetesimals \citet{bot07} or long-term erosion of the asteroid belt \citep[][; assuming an early planetary migration] {gom05, min10}. 

Assuming bombardment by Mars-crosser population shown in Fig. \ref{tail}, Nectarian period would have lasted from about 4.2~Gya (when the number of Mars-crossers has declined by about 20), until the Imbrium impact at 3.87-3.85 Gya. In section 2, we argue that basin magnetization signatures require a substantial gap between the formation of most Nectarian basins and Imbrian ones. If the last pre-gap $D>$900~km basin was Serenitatis at about 4.0 Gya (see above), this would leave about 150 Myr for the decay of the lunar magnetic field (which would have already been relatively weak at the time of Serenitatis impact). This timing of Nectarian basin formation is consistent with the existence of lunar dynamo at 4.2 Gya \citep{gar09}, and the hypothesis of a nutation-driven dynamo \citep{dwy05, mey11}. \citet{mey11} find that the strong dynamo was most likely when the Moon was at a distance of 34 Earth radii from Earth, which corresponds to about 250 Myr after its formation if early Earth's tidal quality factor Q was about 100 \citep[without strong constraints, this value is certainly allowed;][]{bil99}. While detailed chronology (and the issues with dating of Serenitatis) will be discussed in Section 8, we will conclude here that the Mars-crossers are the likely candidates for Nectarian impactors, as they they plausibly could have had the right size-distribution, they survived later than other primordial impactor populations and they can satisfy the chronological constrains based on basin remanent magnetization.

\section{Mutual Collisions and the Imbrian "Cataclysm"}

Primordial Mars-crosser population was likely insufficient to produce all of Imbrian basins and craters. In light of conclusions of Sections 2 and 3, this is not surprising, as there we find that there was a change in the impactor population between the Nectarian system and the Imbrian system. While the Nectarian impactors can plausibly be primordial (and we made a case for Mars-crossers), at least some Imbrian impactors had to be released later, in an event that we could call a "minimal lunar cataclysm" or "Imbrian cataclysm". 

What do we know about Imbrian impactors? \citet{cuk10} argue that their size-distribution was not asteroidal and therefore they cannot be a product of a gravitational destabilization of an asteroid-like reservoir. This conclusion is based in part on the findings of \citet{wil78}, who found that Imbrian impactors had an excess of small bodies compared to Nectarian ones. Regarding their composition, \citet{and73} thought that the Imbrium impactor was closest in trace element signature to type IVA \citep[cf. ][]{gan72} or IIIA iron meteorites. \citep{jam02} found that pre-cataclysm impactors were chondritic, while those at 3.9 Gya were a mix of rarer types, including type IAB irons and pallasites. \citet{puc08} agrees with the non-chondritic (possibly iron meteorite) identification for the Imbrium impactor \citep[as does][]{kor87}, while finding more chondritic composition for some other lunar impactors. The rate of impacts after Imbrium and Orientale impacts dropped rather fast \citep{sto00, ccg07}, and the rate of this decay is consistent with NEAs which are Earth-crossers, Mars crossers or a mixture of those two \citet{cuk10}, but not with a very long-lived population (with a half-life $>100$~Myr). 

An important factor to consider here are mutual collisions between Mars-crossers. We postulate that most of their mass was locked in Vesta-sized bodies. A disruption of one of them may be able to both increase the probability of large basin formation and modify the size-distribution of smaller craters by injecting small impactors with a steep size-distribution into the Mars-crosser region. If the Imbrium impactor was about 60-km across, a thoroughly-disrupted Vesta sized body could produce hundreds of such impactors, enough for one of them to strike the Moon. Since there are about 30 bodies capable of disrupting Vesta at 10 km/s in our reference population (i,.e. with D$>$ 90 km), and the population at 550 Myr (one Mars-crosser half-life before the Imbrium impact) was about 8 times the reference one (with 2\% of 400 still surviving), and the mutual collision probability in our simulation at 600 Myr was $1.3 \times 10^{-17}$, we get that one of "Vestas" should be disrupted every 400 Myr. As our population decays with a half-life of about 200 Myr at this point, probability of such disruption at 550 Myr or later is about 25 \%. Such an event is not improbable, and naturally explains both the new impactor size distribution and increased impact rates during the formation of the Imbrian basins. 


How large would the Wetherill's object need to provide the Imbrium and Orientale impactors? Since most mass is in the Imbrium impactor and its energy was of the order of $10^{26}$~J, and assuming $v=20$~km/s, Imbrian impactor may have had a mass of about $5 \times 10^{17}$~kg, and the total mass of the Wetherill planetesimal would need to be about $2 \times 10^{20}$~kg or larger, assuming Mars-crosser-like impact probabilities. The mass of the Orientale impactor does not change this estimate much (especially as the largest impacts suffer from small-number statistics). Smaller impactors can safely be ignored, as about 700 Imbrian impactors forming craters with $D \ge 20$~km likely carried only about $10^{16}$~kg. Imbrian craters in the 12-100~km range have the cumulative size distribution close to $D^{-2}$ \citep{cuk10}, implying only a few times larger mass in impactors making 100~km craters. Note that the extrapolation of this size distribution would predict several 300~km basins and no $D>900$~basins, implying that basin-forming impactors did not follow this power law, and there may have been a "bump" at the size of Imbrium impactor (Fig. \ref{sfd}). If the Imbrium impactor was indeed metallic, then it seems that even the core of the Wetherill's object was disrupted. Therefore, if we assume that most of the mass of the Wetherill's object was in the Imbrium impactor-sized fragments, we arrive to a total mass estimate close to the mass of Vesta (which is $3 \times 10^{20}$~kg).

\section{Connection to Mesosiderites?}

While the the disruption of the Wetherill's object would explain the peculiarities of the lunar bombardment in the Imbrian era, additional evidence of its late disruption would be invaluable. Apart from the lunar cratering record, the only other current window in the absolute chronology of small-body populations is offered by meteorites. We have already described the constrains on the Wetherill's object from the Imbrian cataclysm: it should be about the size of Vesta ($D\simeq 500$~km), formed in an environment similar to Vesta's, differentiated, and disrupted at about 3.9 Gya. We propose that the mesosiderite group of stony-iron meteorites satisfies all of these constrains. In addition to the similarity in inferred properties between mesosiderite parent body (MPB) and the Wetherill's object, our hypothesis would also explain the puzzling absence of the MPB from the current asteroid belt.

The mesosiderites are made of about equal mix of iron and silicates \citep{hut04}. The silicate component resembles eucrites, and is generally thought to have originally formed close to the surface of  a large differentiated asteroid \citep{sco01}. The metallic component is similar to type IIIAB iron meteorites, and has presumably originated in a core of a differentiated body. Available evidence points to the iron-silicate mix being established early in the history of the solar system, presumably in a dramatic disruption and re-accretion event \citep{sco01}. Apart from their unusual composition, cooling histories of mesosiderites are also peculiar. Metallic component appears to have cooled exceptionally slowly, $0.02$~K/Myr \citep{haa96}. This has been taken as a sign that mesosiderites were buried deep within a large ($D \leq 400$~km) asteroid for hundreds of Myr \citep{pet92}. Ar-Ar dating of the silicate component gives ages that cluster around 3.94$\pm$0.1 Gya \citep{rub93, bog98}, but with indications that at least part of the Ar-Ar ages comes from cooling rather than a discrete event. Still, slow cooling inferred from the metallic component ($0.02$ K/Myr) may be too slow to explain the the argon closure for the silicates at 3.94 Gya \citep{bog98}. The cooling of silicate portion at 3.94~Gya may have been greatly accelerated by a breakup event, explaining the mismatch in cooling rates. However, the lack of strong shocking in mesosiderites has been used to argue against a collisional disruption \citep{haa96}, and is used as evidence that the late argon dates reflect gradual colling only. We hypothesize that the lack of shocking is due to the already fragmented rubble-pile structure of the body resulting from its ancient disruption in which the differentiated layers were mixed. In any case, it is hard to see how mesosiderites could have been excavated from deep within the body without some kind of collisional disruption.


Disruption of the MPB on an unstable orbit would also naturally explain the absence of large amount of mesosiderite parent body material in the main asteroid belt. Even if the mesosiderite Ar-Ar ages are due to cooling and had nothing to do with disruption of the MPB, mesosiderites had to be somehow  excavated from great depths. Is is hard to see how the MPB would not be completely disrupted in the process. Even if it re-accretes, it would have a large family which should still be observable, due to its unusual composition, in addition to expected orbital clustering. If the mesosiderites are excavated from their parent body later than the 3.8~Gya (by which time the asteroid belt must have reached its present state), there is no dynaical way to disperse or hide such a family. This lack of any family is a serious argument against metal-rich asteroid Psyche being the MPB \citep{dav99}. Relatively small size makes the asteroid Maria an unlikely candidate for the MPB, despite some spectroscopic similarities to mesosiderites \citep{fie11}. More broadly, if the mesosiderites were exavated from deep within the MPB, large amount of olivine mantle material must have been excavated, too. While the mesosiderite breccias managed to avoid containing large amounts of olivine \citep[possibly due to the elements of the molten core preferentially attaching to relatively cool crustal materials;][]{sco01}, it is all but impossible to have the whole re-accumulated MPB be olivine-free. It is widely agreed that the mesosiderites cannot be typical of the their parent body's overall composition \citep{sco01}, so we would expect the surviving MPB to be a giant, re-accumulated, olivine-rich asteroid, which is clearly not observed. This need to destroy and re-accumulate the MPB at least once and possibly twice makes Vesta the unlikely candidate for MPB, despite the similarity in the isotopic signatures of MPB and HED meteorites \citep{hut04}. The isotopic similarity between some IIIAB irons, HED meteorites and mesosiderites more likely implies some amount of uniformity in isotopic composition between large differentiated asteroids that were abundant around 2 AU. In the past, main-group pallasites were also considered isotopically indistinguishable from HEDs and mesosiderites, but better techniques can now distinguish their isotopic makeup from that of the other two groups \citep{gre06}. We note that no known iron meteorites have very slow cooling signatures like mesosiderites, implying that no pure iron fragments from the MPB core survived to the present. This is was either due to chance or, more likely, the core may have been broken-up into a smaller number of large objects, none of which survived scattering by the planets.

The fact that the mesosiderite parent body was thoroughly disrupted and reaccreted early in its history, likely through a giant low-velocity collision, is in line with our concept of a very massive primordial Mars-crosser population, if not with out initial conditions (where $e=0.2$ for all particles). As we noted before, our initial conditions are artificial and it is likely that our Mars-crossers-to-be were originally placed on low-eccentricity orbits just beyond that of Mars. In the massive population we are imagining, Vesta-sized bodies should have been disrupted about once on average, most likely early on. However, only if the population was "cold" would Vesta-sized objects be likely to reaccrete (producing the mesosiderite mixture), and survive late enough to become the cause of the lunar cataclysm. The dynamical heating of this population by Mars may have involved migration \citep{min11} and it is clear that our test-particle simulations cannot be taken literally in the first few hundred Myr of the Solar System's history. However, they prove that after being dynamically heated, Mars-crossers naturally leave a Hungaria-like tail.

\section{Survival of Mesosiderites}

A collisional breakup of a Mars-crossing MPB solves the problem of its absence from the current asteroid belt. However, a new problem appears, regarding the survival of any of its fragments. If the precursors of mesosiderites were planet-crossing, it is surprising that any of them are still around. It is clearly impossible that mesosiderites could survive almost 4 Gyr in near-Earth space \citep{bot07}. However, our own integrations have shown that on the order of 1\% of Mars-crossers become trapped in the Hungaria region, decaying very slowly. Assuming that Hungarias were depleted by two orders of magnitude since the fragments' implantation (as about 6.5 of the 600-Myr Hungaria lifetimes have passed since LHB), we find that about 0.01\% of the MPB's fragments should have survived among Hungarias, based on dynamical considerations alone. These would include about 100 km-sized bodies (of one million original ones) or about one 10-km body. Collisional evolution of Hungarias complicates this calculation, as km-sized cannot survive to present but new ones could have been made from the larger bodies that were disrupted before they had chance to escape. While the mesosiderites could not have been favored by the gravitational dynamics, it is reasonable to assume that the mesosiderites would be selected for by collisional grinding over olivine fragments of the mantle. So as long as mesosiderites are mechanically stronger, and one of the D$>$10 km fragments that evolved into Hungarias contained a few percent of the mesosiderite material, the survival of mesosiderites is reasonable.

There is an additional effect that can account for the survival of mesosiderites. If the eccentricity of Mars was higher than the long-term average during the MPB disruption, and has subsequently decreased, a larger fraction of MPB fragments could have been implanted into the Hungaria region. If the eccentricity of Mars was increasing prior to the lunar cataclysm (i.e. went through a maximum around 4 Gya), it could explain both the relatively late timing of the cataclysm and help the survival of mesosiderites. An increasing eccentricity of Mars could somewhat change the decay rate of impactors showing Fig 4. If the eccentricity of Mars decreases after some of the MPB fragments have been injected into the Hungaria region, large fraction of those fragments could be basically stable on the age of the Solar System, rather than decaying with a 600~Myr half-life expected from the "average" secular histories. In any case, MPB fragments are unlikely to make up more than a small minority of all Hungaria asteroids. This estimate is based on the fact that the cataclysm should have produced about three times as many as many bodies as there were left from formation at this epoch, but that only about few percent of these MPB fragments could have been expected to decouple from Mars and become proto-Hungarias (which were already a large part of the background population).  In addition to the varying eccentricity of Mars, non-gravitational forces could have also contributed to the survival of MPB fragments \citep[ through the "adoption" mechanism of ][]{mce10}. However, this mechanism works best for km-sized pieces and does not seem efficient enough to dominate over purely gravitational processes.

We have assumed that the long-term refuge of mesosiderites would be the Hungaria region, as we have previously found that it is a natural storage for remnants of planetary accretion. However, there are other possibilities for long-term survival of small sturdy fragments. \citet{bot07} indicate that the high-$i$ orbits in the innermost asteroid belt may serve as long-term refuge to a very small fraction of bodies from the inner Solar System. As about a million small km-sized bodies would have been injected into Mars-crossing orbits by the disruption of the Wetherill's object, some of them may have found very low-probability dynamical pathways to long-term stability. One remote possibility is that some of the martian Trojans may have been captured at this time, as they may be long-lived but not strictly stable \citep{con05, sch05}. One advantage the martian co-orbitals would have over Hungarias as long-term survivors is a much more benign collisional environment on planet-crossing orbits. While we think that Hungarias are the most likely current source of mesosiderites, further work on the dynamics and surface composition of Mars Trojans \citep{riv07} should be able to test the possibility that at least some of them originated in the Imbrian cataclysm.

\section{Discussion}

Here we will detail a new scenario of the chronology and sources of early lunar bombardment. This scenario is tentative and will need further improvements if it is to explain the full complexity of the lunar record. However, we believe that by proposing a number of testable predictions we can help advance lunar science.

On the basis of magnetic and cratering data, we conclude that there were two populations of ancient lunar impactors, a pre-Imbrian one and an Imbrian one. These groups are largely equivalent to Populations I and II from \citet{str05}, although there is some disagreement about the nature of Population II. Here we find that the most likely source of Population I were the primordial Mars crossers, which would have decayed with the half-life of 80 Myr and may have been able in producing $D > 900$~km basins until 4.0~Gya. The size-distribution of the Population I impactors was very similar to present-day main belt asteroids (MBAs), but there is no reason to think that other (now extinct) inner solar system small body populations did not have the same SFD. It is reasonable to expect that MBAs, Mars-crossers, the "E-belt" \citep{cuk08, bot10} and the leftovers from Earth's formation \citep{bot07} would all have had similar size-distributions.


The above discussion implies that the size-distribution and composition of Population I impactors are unlikely to be specific to their source region. However, chronology of their impacts should be indicative of their dynamical evolution. If the Nectaris basin is older than about 4.2 Gyr, it could plausibly be the product of a primordial impactor population, as proposed here. If, on the other hand, it is younger than 4.1 Gyr ago, it is likely to be a part of a cataclysm-type event \citep{bot10}. The absolute age of Nectaris basin may be impossible to determine from the existing Apollo 16 samples, which appear dominated by Imbrium ejecta \cite{kor87, has98, war03, nor10}. The 3.9~Gya Apollo 17 ages attributed to Serenitatis may actually reflect the formation of Imbrium, with Serenitatis being much older \citep{has98, spu11}.  Unfortunately, the factors that complicate dating of the nearside Nectarian basins may similarly affect future samples from other locations on the Moon. Moscoviense may be the only lower Nectarian basin sufficiently distant from both Imbrium and Orientale impacts that it was likely not contaminated much by their ejecta, and may be (in theory) the best candidate for sample return among Nectarian basins. Dating of the largest and oldest South Pole-Aitken (SPA) basin would put a upper limit on the age of lunar surface features and confirm or rule out the total resurfacing scenario \citep{ter74}, but would not give us any information about the chronology of much later impactors. An age of about 4.4~Gyr for SPA would allow both "medium" \citep{bot10} and minimal (as proposed here) versions of the lunar cataclysm, so at least one Nectarian basin would need to be dated in order to distinguish between these two tentative chronologies. 

In contrast, the timing of (lower Imbrian) Population II impactors is well established, as the Imbrium basin formed at 3.85-3.87 Gya, and the most Imbrian craters formed shortly afterwards. Since the size distribution of Imbrian impactors appears to be different from that of Nectarian ones \citep{wil78, cuk10}, and different from asteroids, their source is unknown. We propose a collisional disruption of a Vesta-sized Mars-crosser as their source \citep{wet75} on the basis of theoretical considerations, tentative composition of the Imbrium impactor \citep{puc08}, and possible connection to mesosiderites.

A minimal cataclysm from a Mars-crossing collisional disruption also affected Mars, but the number of superposed craters indicate that the largest impact features on Mars like Argyre and Hellas are likely to predate the lunar cataclysm \citep{tan86, wer08}. In any case, amount of mass in 200 km bodies necessary to form these basins is not consistent with the breakup of one Vesta-sized body. Our hypothesis predicts that none of the largest basins formed at Mars at 3.85 Gya, and, in general, implies much older ages for Noachian features than the models that assume a larger-scale bombardment at 3.85 Gya \citep{wer08}. Despite the limited extent of the spike at 3.9 Gya, Mars would have experienced substantial bombardment prior to 3.6 Gya due to Mars-crossers and this bombardment was the likely the source of shocking found in the ancient martian meteorite ALH84001 \citep{ash96}.
 
Shock ages of HED meteorites and H-chondrites are also sometimes used to constrain the early Solar System bombardment \citep{bog03, swi09}. Interestingly, both HED meteorites and H-chondrites record a much longer bombardment episode than the Moon, making it hard to explain both records with the same impactor population \citep{har03}. We think that it is significant that the HED and H-chondrite parent bodies \citep[Vesta, and, most likely, Hebe ]{mcs10, gaf98, bot10b}  are both located in the inner main belt and would be likely to suffer significant number of collisions with late surviving Mars-crossers/proto-Hungarias. The shock resetting of meteorite Ar-Ar ages appears to be dependent both on the size of the body, parameters of impacts and the family history \citep{nes09}. Therefore, while the number of impacts by Mars-crossers onto Vesta and Hebe from 3.5 to 4.1 Gya did not dwarf their collisions with other asteroids since then (H-chondrites also show more recent Ar-Ar ages, while HEDs do not), it is likely that the shock ages of HED meteorites and H-chondrites do have a connection to Mars-crossers. Even more than meteorite shock ages, impact events recorded in lunar meteorites \citep{coh00, coh05} seem at odds the lunar impact history inferred from Apollo samples \citep{har03, har07}. Events postdating 3 Gya are recorded in lunar meteorites, long after the heavy bombardment ended. Our hypothesis does not offer an explanation for this discrepancy, and we suspect that the solution lies in the physics of lunar regolith rather than the dynamics of impactors.

How can our hypotheses be tested? First, new crater counts (down to about $D=8$~km) on all non-magnetized basins (Imbrium, Orientale, Schr\" odinger and Hertzsprung) would confirm or falsify the existence of a non-asteroidal Imbrian size distribution, without having to resort to stratigraphic or morphological identification of craters. Otherwise, the study of mesosiderites and Hungarias could probably test their relationship with the lunar cataclysm faster than new lunar samples can be collected. Our hypothesis predicts that the mesosiderite parent body (or bodies) are likely located among Hungarias (although innermost main belt and Mars Trojans are also plausible refuges).  Remote sensing of asteroid composition has advanced considerably in the last few decades, and it is possible that we may be able to identify the intermediate mesosiderite parent body, if it is big and bright enough (small Hungarias are brighter than small MBAs). Radar observations may also in principle be able to break the degeneracy among X-type asteroids among Hungarias \citep{ock10}. Additionally, if the Dawn spacecraft identifies a mesosiderite unit exposed on Vesta (or if mesosiderites are traced to any other large asteroid deep in the main belt), our hypothesis about the connection between the mesosiderites and the cataclysm would be falsified. While the collisional disruption hypothesis can be formulated independently from mesosiderites \citep{wet75}, the case becomes weaker once mesosiderites are excluded from consideration.


\section{Summary}

In this work we propose several hypotheses regarding the lunar cratering record and the lunar cataclysm:

1. We interpret the magnetization signatures of lunar basins to indicate that there was a chronological gap between the formation of most Nectarian basins and Imbrium basin. This interpretation hinges on 3.9-Gya Apollo sample dates attributed to Nectaris and Serenitatis actually reflecting the age of Imbrium basin\citep{has98}. 

2. Size-frequency distributions of lunar craters agree with a two-stage bombardment, with most Nectarian craters having an asteroid like size distribution, and the Imbrian ones containing more small impactors \citep{cuk10}.

3. We find that primordial Mars-crossers (with original $a< 2$~AU) should have a 80~Myr half-life and could have produced more observable lunar craters than any other primordial impactor population.

4. We propose that Nectarian and older "Population I" craters on the Moon \citep{str05} were overwhelmingly produced by Mars-crossers. This dynamical model predicts South Pole-Aitken basin being about 4.35 Gyr old, Nectaris about 4.25-4.2 Gyr old and the last giant Nectarian basin (possibly Serenitatis or Crisium) being about 4.05-4.0~Gyr old.

5. A long-lived (but still eroding) remnant of this population is still present as the Hungaria asteroid group. This makes the Hungarias primordial, but not truly stable over the age of the Solar System.

6. Imbrian basins (and possibly Hertzsprung) formed about 3.9-3.8~Gya, in what is a separate cratering spike. We propose that this spike was caused by collisional disruption of a Vesta-sized body ("Wetherill's object") on a Mars-crossing orbit.

7. Wetherill's object was a late-surviving member of the same Mars-crossing population that cratered lunar highlands.

8. Wetherill's object was also the original parent body of mesosiderite meteorites.

9. Intermediate parent bodies of mesosiderites diffused into the Hungaria region from Mars-crossing orbits. They have since suffered dynamical and collisional attrition, with the latter favoring metal-rich mesosiderites over more numerous silicate fragments of the mantle.

Hopefully, future studies of the Moon, asteroids and meteorites will provide data against which these hypotheses can be tested. 

\bigskip

\section*{Acknowledgments}

The author thanks Catherine Johnson and Surdas Mohit for sharing their unpublished data, Zo\" e Leinhardt, Ben Weiss, Bill Bottke, David Minton and Mark Wieczorek for useful discussions, and Sarah Stewart-Mukhopadhyay and Brett Gladman for their help and support during the whole LHB project. Insightful reviews by Bill Bottke and Lon Hood greatly improved the paper. Numerical integrations were conducted on Harvard University's "Odyssey" computer cluster. This work was done while M. {\' C}. was Clay Postdoctoral Fellow at the Smithsonian Astrophysical Observatory. This paper is dedicated to the memory of Brian Marsden.

\bigskip

\newpage

\bibliographystyle{}

\newpage

\begin{table}[htdp]
\begin{center}
\begin{tabular}{|c|c|c|}
\hline
Reference and & W78 Imbrian& W78\\
crater population & \& younger & Nectarian \\
\hline
W78 Imbrian and younger & 100\% & {\bf $<$ 0.1\%} \\
W78 Imbrian craters only & $>$ 10\% & {\bf $<$ 0.1 \%} \\
S05 class 1 craters ($D>$22 km) & $>$ 10 \% & $<$ 2.5 \% \\
S05 highland craters ($D>$22 km) & {\bf $<$ 0.5\%}& $>$ 10 \% \\
W87 craters on HIOS basins & 2.5-5\% & $<$ {\bf 1 \%} \\

\hline
\end{tabular}
\end{center}
\caption{Results of Kolmogorov-Smirnov two-population tests for different crater populations discussed in this paper. Abbreviations used in the table: S05 - \citet{str05}; W78 - \citet{wil78}; W87 - \citet{wil87}.}     
\label{KStest}
\end{table}

\begin{table}[htdp]
	\begin{tabular}{|l|c|c|c|}
            \hline
            Planet / & 2$^{\rm nd}$ generation & Earth-crosser & Mars-crosser \\
	    Outcome & Mars-Crossers  & break-up & break-up \\
            \hline
            Sun &  70.8 \% &  81.6 \% & 78.7 \% \\
            Mercury & 3.3 & 1.8 \% & 0.9 \%\\
	    Venus & 10.1 \% & 5.6 \% & 6.5 \% \\
            Earth & 7.6 \% & 6.4 \% & 5.3 \% \\
           Mars & 2.5 \% & 0.5 \% & 2.3 \% \\
	   Outer/Ejected & 5.8 \% & 4.2 \% &  6.4 \% \\
         \hline
        \end{tabular}
	\caption{\baselineskip 24pt Fates of different populations of test particles. Particles were removed after hitting the Sun, approaching within one radius from a planet, or reaching the heliocentric distance of 100 AU. First column plots outcomes for second generation mars-crossing particles that were removed from simulation featured in Figs 1-3 between 300 Myr and 1000 Myr. The other two columns list the end-fates of fragments from typical disruptions on high-inclination Earth-grazing and Mars (but not Earth)-crossing orbits. }
\label{fates}
\end{table}

\newpage 
\pagestyle{empty}

\newcounter{subfigure}
\renewcommand{\thefigure}{\arabic{figure}\alph{subfigure}}
\setcounter{subfigure}{1}

\begin{figure*}
\includegraphics[angle=270, scale=0.7]{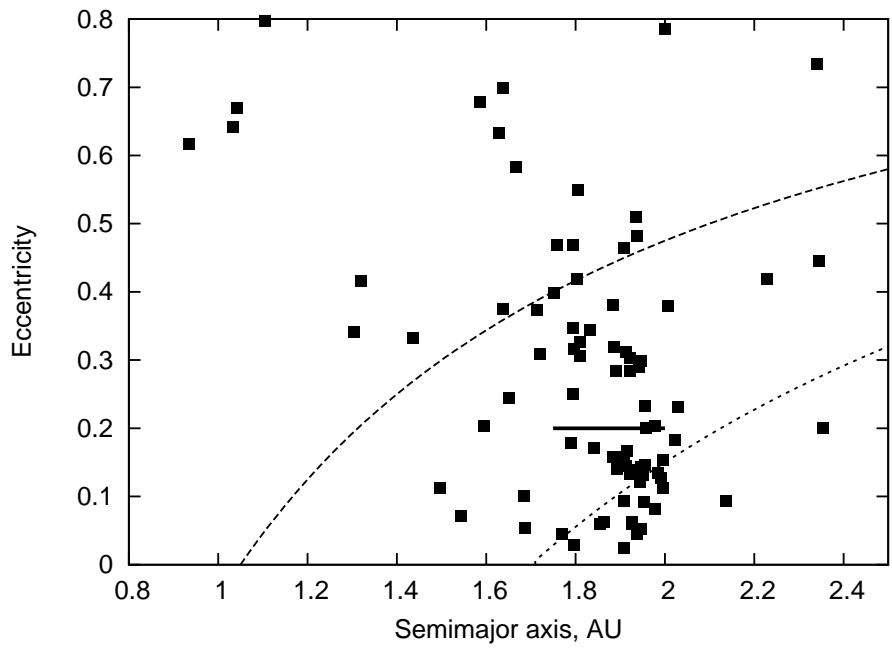}  
\caption{Semimajor axis-eccentricity distribution of the remaining first generation test particles at 300~Myr.} 
\label{300myr_ae}
\end{figure*}

\addtocounter{figure}{-1}
\addtocounter{subfigure}{1}

\begin{figure*}
\includegraphics[angle=270, scale=0.7]{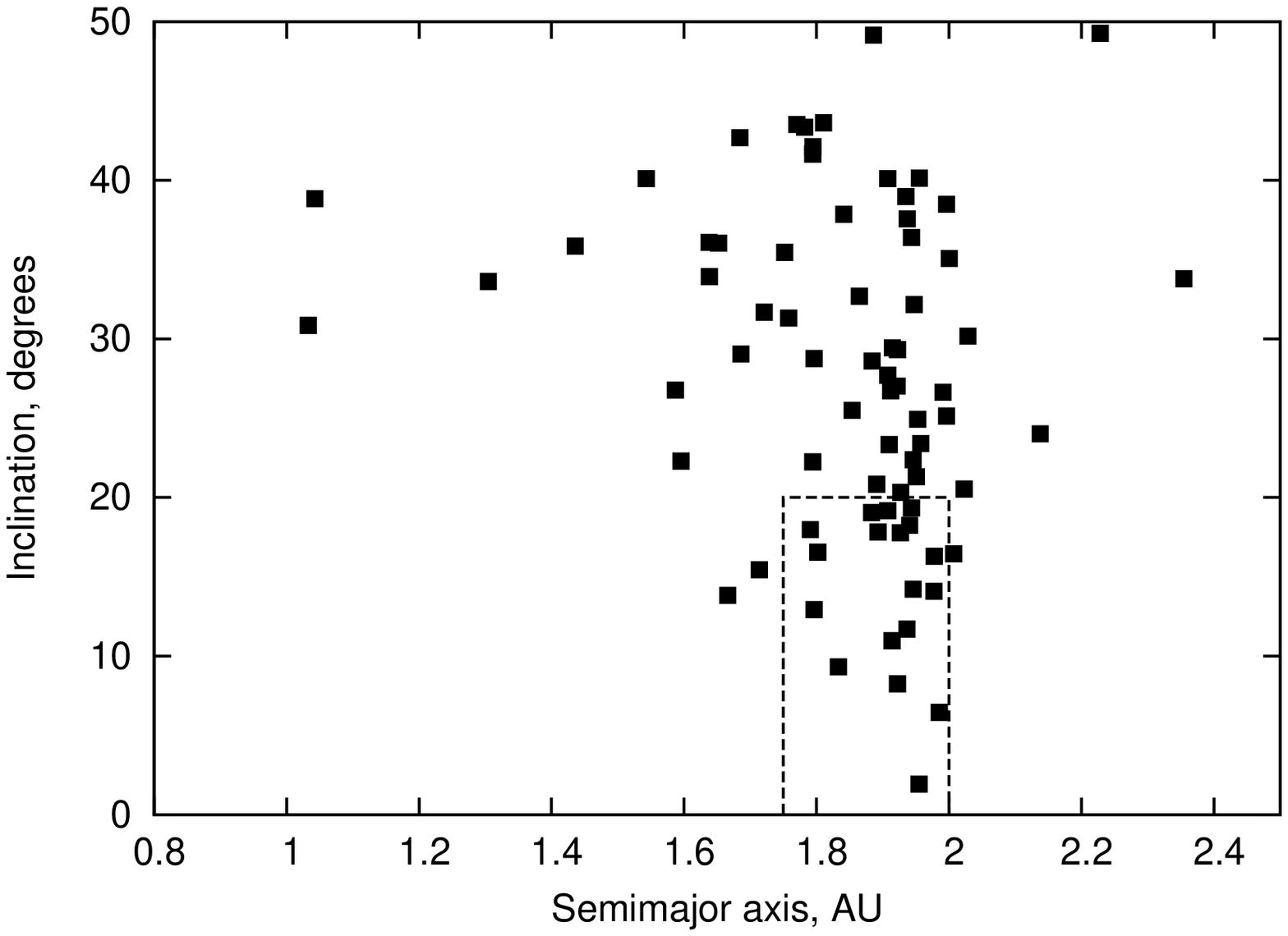}  
\caption{Semimajor axis-inclination distribtion of the remaining first generation test particles at 300~Myr.} 
\label{300myr_ai}
\end{figure*}

\setcounter{subfigure}{1}

\begin{figure*}
\includegraphics[angle=270, scale=0.7]{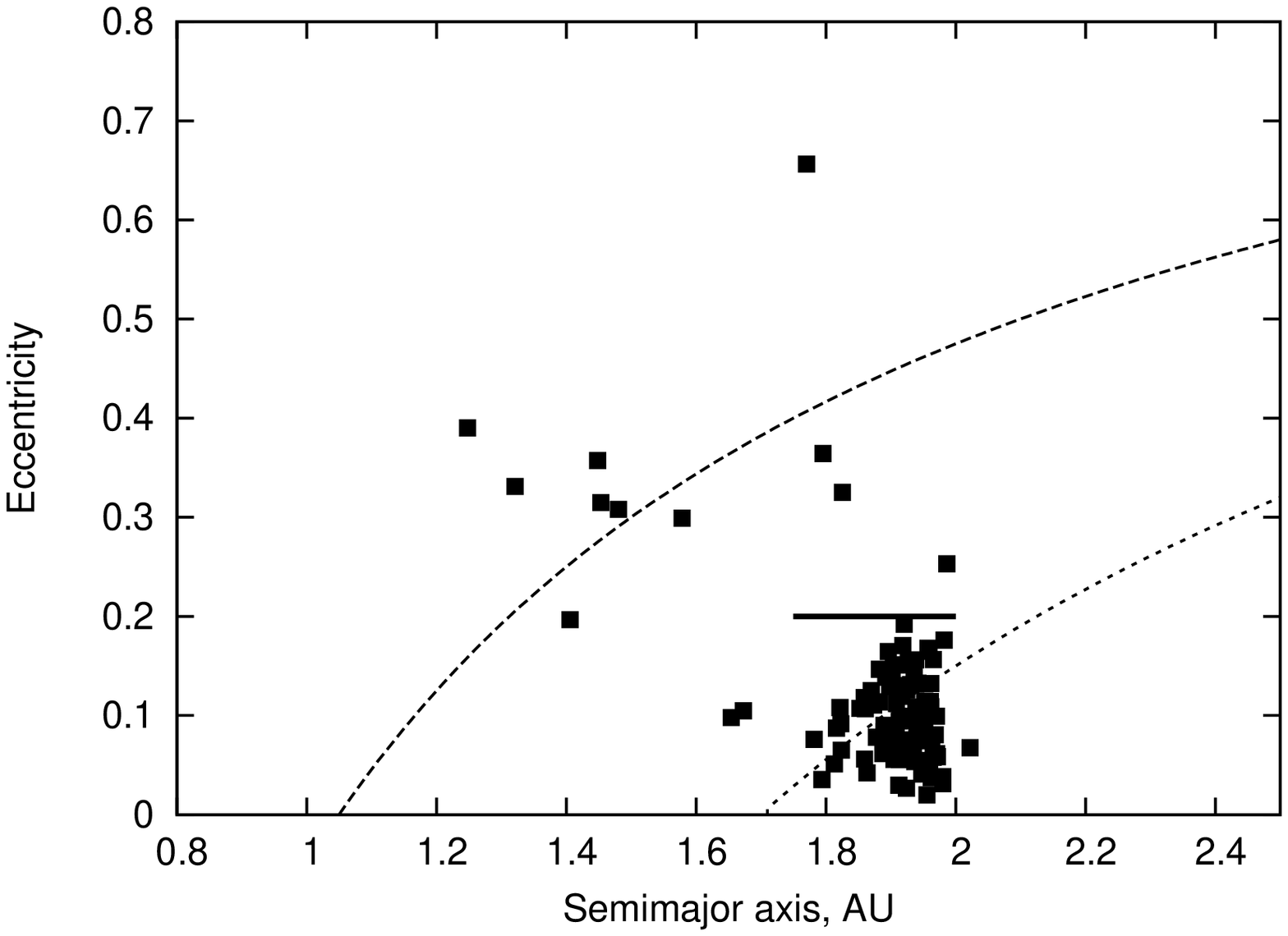}  
\caption{Semimajor axis-eccentricity distribtion of the remaining second generation test particles at 1~Gyr.} 
\label{1gyr_ae}
\end{figure*}

\addtocounter{figure}{-1}
\addtocounter{subfigure}{1}

\begin{figure*}
\includegraphics[angle=270, scale=0.7]{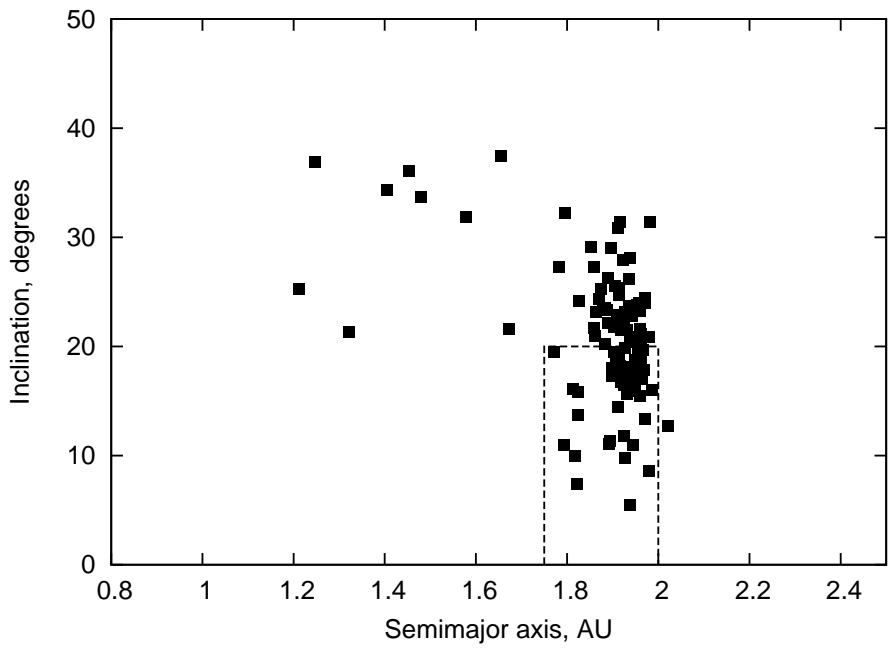}  
\caption{Semimajor axis-inclination distribtion of the remaining second generation test particles at 1~Gyr.} 
\label{1gyr_ai}
\end{figure*}

\renewcommand{\thefigure}{\arabic{figure}}





\begin{figure*}
\includegraphics[angle=270, scale=.7]{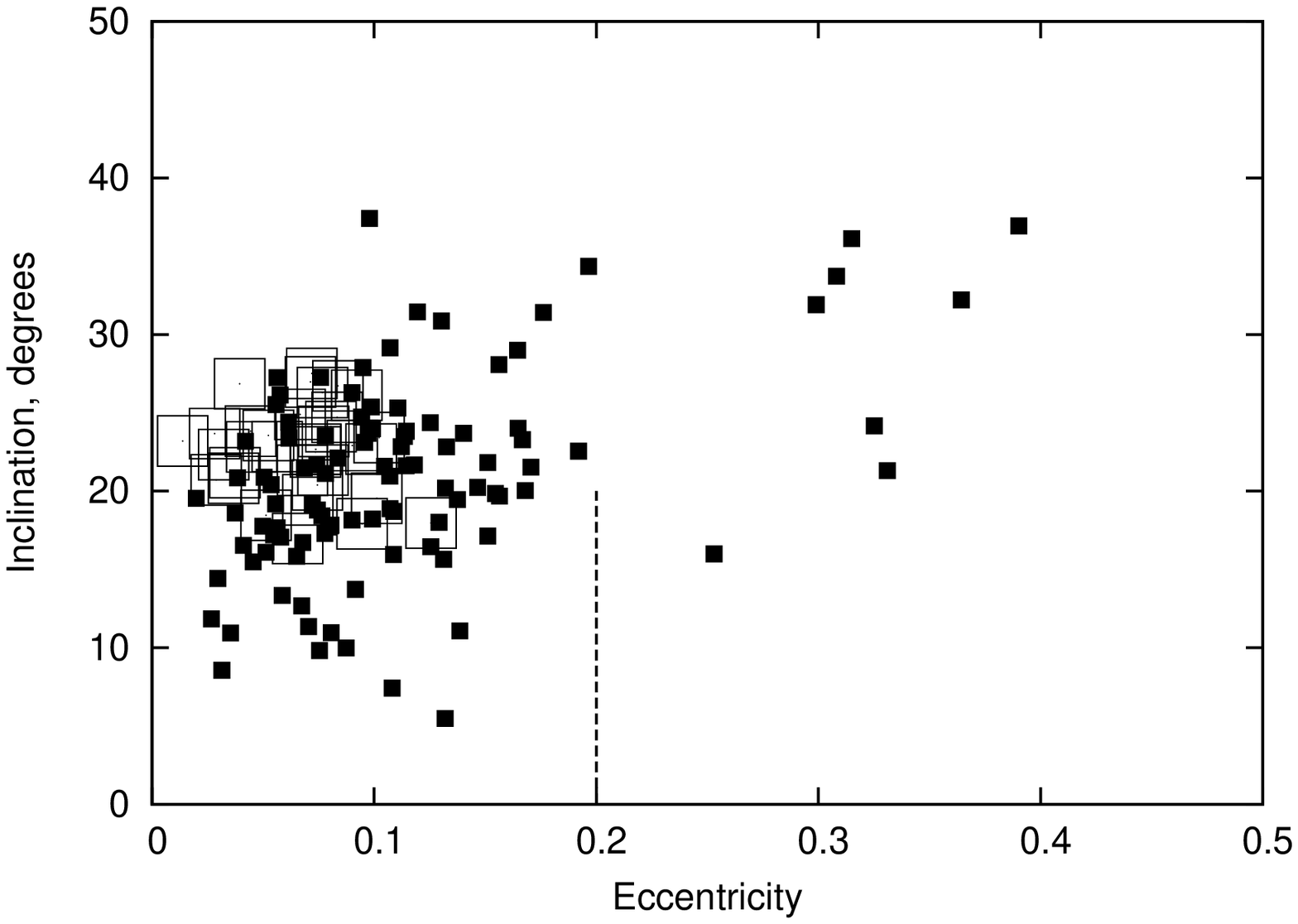}  
\caption{Comparison between the eccentricity-inclination distribution of test particles at 1000 Myr from Figure 2 and Hungaria asteroids (filled squares) and 35 brightest Hungarias (open squares). All first-generation particles had initial conditions the dashed line ($0 < i < 20^{\circ}$, $e=0.2$).}
\label{hungaria}
\end{figure*}

\begin{figure*}
\includegraphics[angle=270, scale=.7]{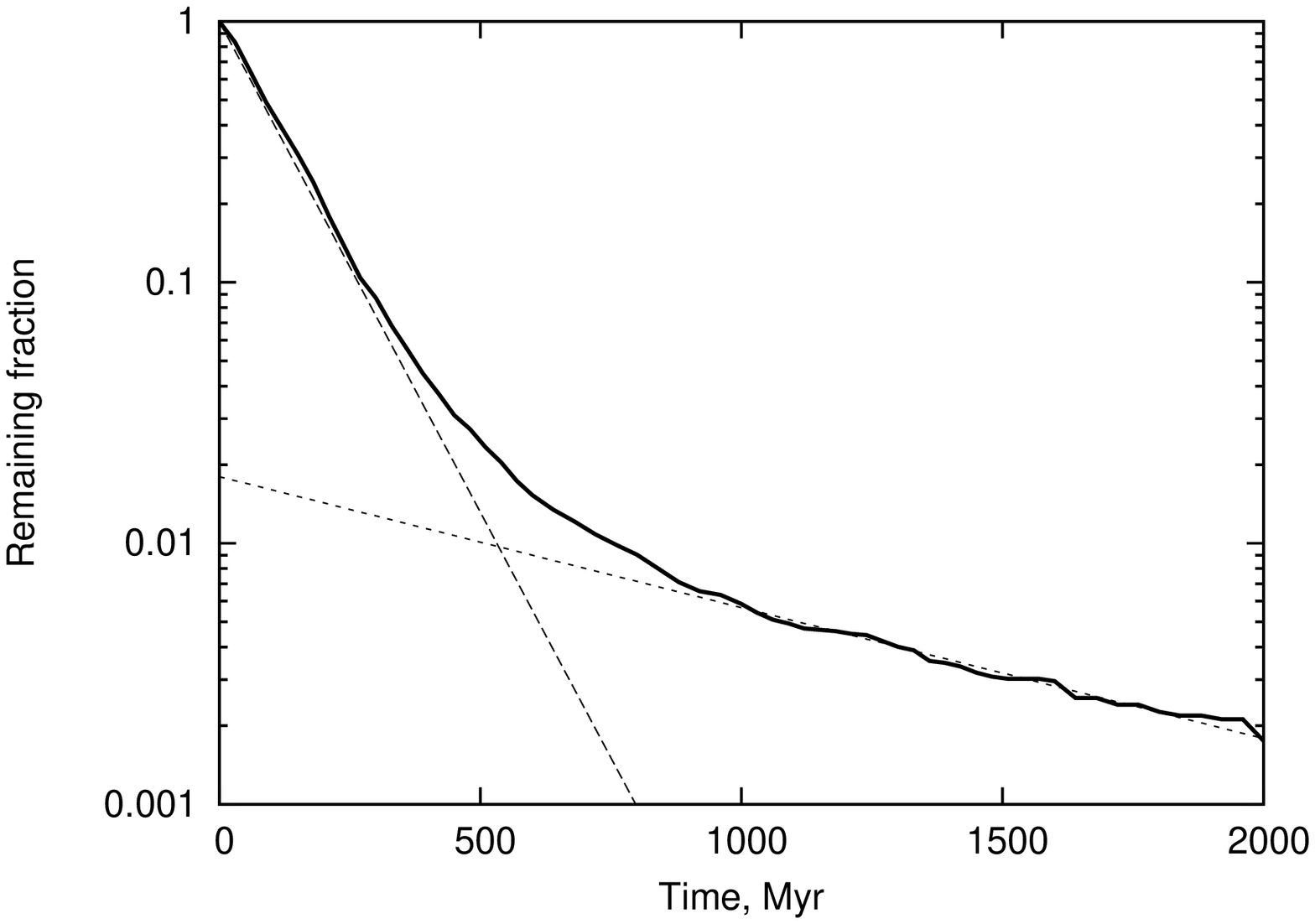}  
\caption{Remaining Fraction of our Mars-crosser population over 2 Gyr (solid). The straight dashed lines trace the initial 80-Myr half-life decay of Mars-crossers and the later 600-Myr half-life decay of proto-Hungarias.}
\label{tail}
\end{figure*}

\begin{figure*}
\includegraphics[angle=270, scale=.7]{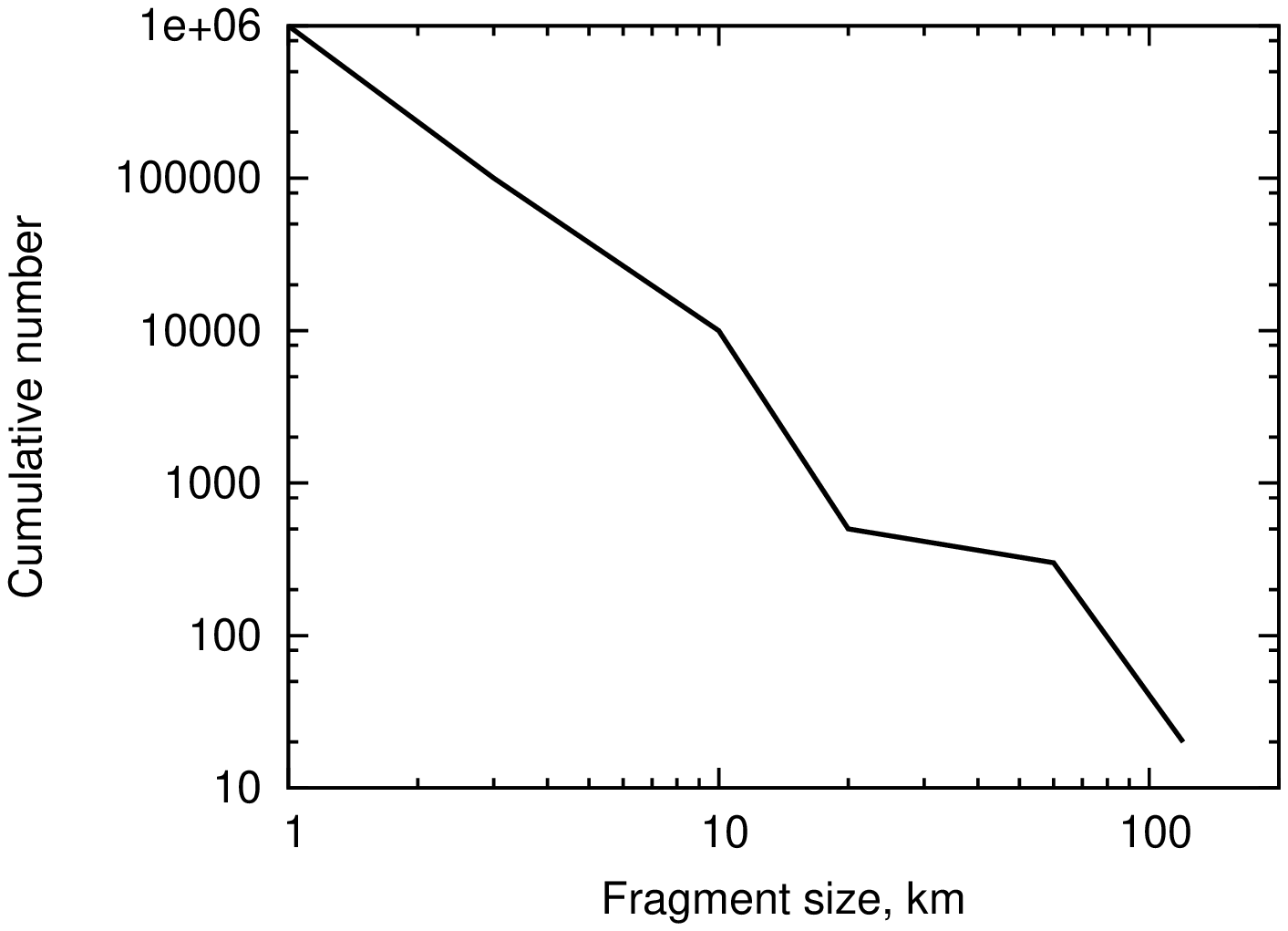}  
\caption{Rough estimate of the size distribution of fragments from the postulated Wetherill's object disruption. We assume that much of the mass is in Imbrium impactor-sized pieces (about 60 km in diameter), and with fragments below about 10 km having the cumulative distribution close to $N(>D) \sim D^{-2}$. Note that this slope is determined from the "population 2" crater SFD on the Moon and may not reflect impactor distribution at any one point of time. Since the fragments would be undergoing constant collisional evolution (and "dilution" by the backround Mars-crossers) between the parent body disruption and lunar impact, it is possible that the immediate post-disruption size-disttribution below 10 km may have been steeper. A substantial part of the mass could also have been in few largest pieces but we have no way of inferring their existence.}
\label{sfd}
\end{figure*}

\end{document}